# Carbyne From First Principles: Chain of C Atoms, a Nanorod or a Nanorope?

*Mingjie Liu,[1] Vasilii I. Artyukhov,[1] Hoonkyung Lee,[1,†] Fangbo Xu,[1] and Boris I. Yakobson[1,2,3,*]*

[1]Department of Mechanical Engineering and Materials Science, [2]Department of Chemistry, and [3]Smalley Institute for Nanoscale Science and Technology, Rice University, Houston, Texas 77005, USA

\* e-mail: biy@rice.edu

We report an extensive study of the properties of carbyne using first-principles calculations. We investigate carbyne's mechanical response to tension, bending, and torsion deformations. Under tension, carbyne is about twice as stiff as the stiffest known materials and has an unrivaled specific strength of up to $7.5\times10^7$ N·m/kg, requiring a force of ~10 nN to break a single atomic chain. Carbyne has a fairly large room-temperature persistence length of about 14 nm. Surprisingly, the torsional stiffness of carbyne can be zero but can be 'switched on' by appropriate functional groups at the ends. Further, under appropriate termination, carbyne can be switched into a magnetic-semiconductor state by mechanical twisting. We reconstruct the equivalent continuum-elasticity representation, providing the full set of elastic moduli for carbyne, showing its extreme mechanical performance (*e.g.* a nominal Young's modulus of 32.7 TPa with an effective mechanical thickness of 0.772 Å). We also find an interesting coupling between strain and band gap of carbyne, which is strongly increased under tension, from 3.2 to 4.4 eV under a 10% strain. Finally, we study the performance of carbyne as a nanoscale electrical cable, and estimate its chemical stability against self-aggregation, finding an activation barrier of 0.6 eV for the carbyne–carbyne cross-linking reaction and an equilibrium cross-link density for two parallel carbyne chains of 1 cross-link per 17 C atoms (2.2 nm).

Carbon exists in the form of many allotropes: zero-dimensional $sp^2$ fullerenes,[1] the two- dimensional $sp^2$ honeycomb lattice of graphene[2] (parent to graphite and carbon nanotubes), or three-dimensional $sp^3$ crystals—diamond and lonsdaleite. Each allotrope has notably different electronic and mechanical properties. For instance, graphene has the characteristic semimetal electronic structure with a linear band dispersion[3] and an extraordinarily high electron mobility.[4] In contrast, diamond is a wide-band-gap insulator and one of the hardest natural materials known.

Carbon can also exist in the form of carbyne, an infinite chain of *sp*-hybridized carbon atoms. It has been theoretically predicted that carbyne may be stable at high temperatures (~3000 K).[5] Indications of naturally-formed carbyne were observed in such environments as shock-compressed graphite, interstellar dust, and meteorites.[6–8] The carbyne-ring structure is the ground state for small (up to about 20 atoms) carbon clusters.[9] Experimentally, many different methods of fabrication of finite-length carbon chains have been demonstrated, including gas-phase deposition, epitaxial growth, electrochemical synthesis, or 'pulling' the atomic chains from graphene or carbon nanotubes.[10–17] Recently, chains with length of up to 44 atoms have been chemically synthesized in solution.[18] Many interesting physical applications of carbynes have been proposed theoretically, including nanoelectronic/spintronic devices,[19–23] and hydrogen storage.[24] Moreover, very recently, such complex molecular mechanisms as rotaxanes based on carbyne chains have been synthesized.[25,26] All of these advances make the understanding of mechanical behavior of carbyne more and more important.[27]

Albeit there exists a considerable body of literature on the structure[28,29] and mechanics[30–33] of carbyne, it is very scattered and has thus failed so far to provide a well-rounded perspective. Here we address this shortage and put forward a coherent picture of carbyne mechanics and its interplay with chemical and electronic phenomena.

## Results and discussion

We begin by discussing the structure of carbyne. One textbook structure of one-dimensional chain of C atoms is the cumulene (=C=C=), known to undergo a Peierls transition[34] into the polyyne (–C≡C–) form. Our calculations confirm this instability, showing an energy difference of 2 meV per atom in favor of the polyyne structure. The calculated cohesive energy of polyyne is $\boldsymbol{E_c}$ = **6.99 eV** per atom. It must be noted that pure DFT methods underestimate the band gap, and this results in a much weakened Peierls instability. Higher-level quantum chemistry techniques or exact-exchange hybrid functionals can correct this deficiency,[23,35] however it is not of prime importance for the mechanical properties of carbyne, and thus we do not need to resort to these more costly computational methods.

**Carbyne under tension: the strongest nanowire.** The most basic mechanical property of the carbyne chain is its tensile stiffness, defined as

$$C = \frac{1}{a}\frac{\partial^2 E}{\partial \varepsilon^2}, \qquad (1)$$

where $a$ is the unit cell length (2.565 Å), $E$ is the strain energy per two C atoms, and $\varepsilon$ is the strain. Fitting Eq. (1) to our DFT data with a fourth-order polynomial (**Fig. 1 (a)**) yields a tensile stiffness of $C$ = 95.56 eV/Å in agreement with earlier work.[36] (Note that negative strain values are mainly of academic interest; however, it may be possible to realize some degree of compression in finite-length chains below the onset of Euler

buckling—see next section on bending stiffness,—or with chains confined inside channels, *e.g.*, carbon nanotubes.[37]) A more informative measure from the engineering standpoint is the specific stiffness of the material, which for carbyne works out to be **~10⁹ N·m/kg**. This is a more than twofold improvement over the two stiffest known materials—carbon nanotubes and graphene (4.5×10⁸ N·m/kg[38–40])—and almost threefold, over diamond (3.5×10⁸ N·m/kg[41]).

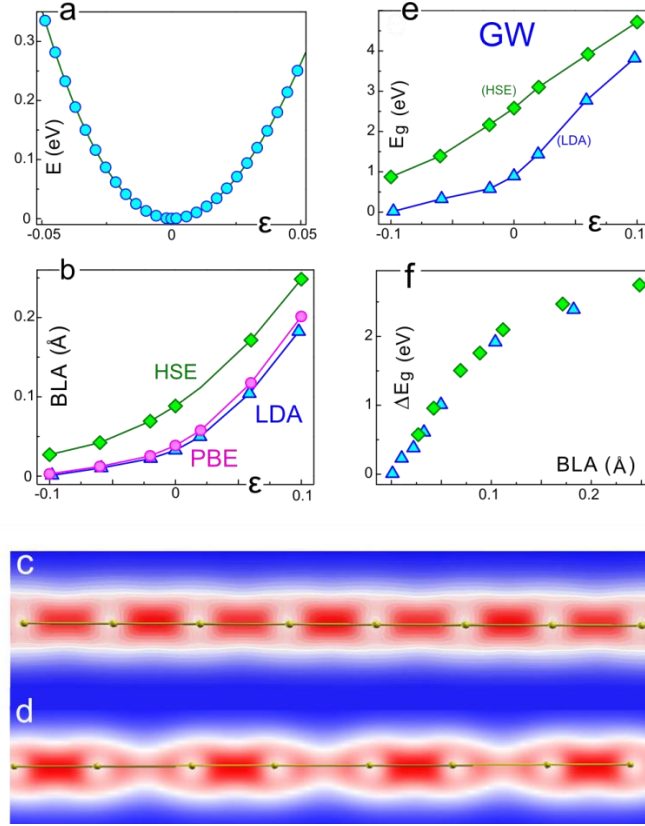

**Figure 1**. Carbyne under tension. (a) DFT calculations of stretching energy per unit cell (two atoms) as a function of strain ε. The bond length alternation (b) and electronic density of carbyne (c) without strain and d) under tension, showing a more pronounced bond alternation in stretched carbyne. (e) GW band gap increase as a function of strain for structures obtained with pure (LDA) and hybrid (HSE) density functionals, the latter predicting a stronger BLA (b). (f) Dependence of GW correction to LDA band gap as a function of bond length alternation.

Another important metric of engineering materials is their specific strength. The tensile strength is generally difficult to predict theoretically, therefore we approach the problem from two complimentary perspectives. One method to evaluate the ideal strength of carbyne is to compute the phonon instability point—the strain at which imaginary frequencies appear in the phonon spectrum.[39] We found that in carbyne, this event occurs at a critical strain of 18–19%, at a pulling force of 11.66 nN. Notably, this coincides with the change from direct to indirect band gap.[42] Another approach is to treat fracture as an activated bond rupture process and compute its activation barrier $E_a(\varepsilon)$ as a function of strain.[43] The computed $E_a(\varepsilon)$ dependence shows the expected decrease with tension, and at a strain of ~9% (9.3 nN) it reaches a characteristic magnitude of 1 eV for which transition state theory predicts an expected lifetime on the order of ~1 day. Thus the estimated breaking force of a carbyne chain is **9.3–11.7 nN** (this is >10× the previous literature report of 0.9 nN,[32] demonstrating the inadequacy of the ReaxFF forcefield used therein for carbyne mechanics). This force translates into a specific strength of **6.0–7.5×10⁷ N·m/kg**, again significantly outperforming every known material including graphene (4.7–5.5×10⁷ N·m/ kg[38,39]), carbon nanotubes (4.3–5.0×10⁷ N·m/ kg[44,45]), and diamond (2.5–6.5×10⁷ N·m/kg[46]). The details of calculations are described in Supplementary Materials.

It is interesting to see what happens to the bond length alternation (BLA) of carbyne under tension.[27,42] Since the strength of Peierls distortion is extremely sensitive to the choice of methods,[47] we performed additional computations using the local density approximation and the hybrid HSE06 functional incorporating exact exchange.[48,49] The most trustworthy method—HSE06—yields an increase of BLA from 0.088 Å in free carbyne to 0.248 Å when $\varepsilon = 10\%$ (**Fig. 1 (b)**). **Fig. 1 (c, d)** shows the electronic densities of carbyne in free-standing state and at a 10% strain. While in the first case (c), the BLA is barely noticeable in the plot, the second (d) shows very clear distinction between the 'single' and 'triple' bonds.

Naturally, this change in the BLA has big ramifications for the electronic properties of carbyne. **Fig. 1 (e)** shows the variation of the band gap with strain computed using many-body perturbation theory ($G_0W_0$) [50] in two settings: the standard LDA+$GW$ scheme and using the more realistic HSE06 geometry (HSE06/LDA+$GW$). The difference between the two plots in **Fig. 1 (e)** is easily understood from the plot of $GW$ correction magnitudes as a function of BLA shown in **Fig. 1 (f)**, where the data points from both sets fall closely on the same line. Clearly, an accurate reproduction of BLA is essential for the computation of band gap. The HSE06/LDA+$GW$ band gap varies almost linearly with strain from 2.58 eV at 0% to 4.71 eV at 10% strain, an 83% change. This extreme sensitivity of band gap to mechanical tension shows carbyne's great prospects for opto-/electromechanical applications. (An independent study reporting similar results[51] came out while the present manuscript was in preparation.) The nature of this effect and its unusual consequences are explored more deeply elsewhere.[23]

**Bending stiffness: nanorod or nanorope?** Given the extreme tensile stiffness of carbyne, it is interesting to investigate how this material performs under bending deformations. The bending stiffness is defined as

$$K = \frac{1}{a}\frac{\partial^2 E}{\partial q^2}, \quad (2)$$

where $a$ is again the unit cell and $q$ is the curvature. To calculate the bending energy of carbyne, we use the model of carbon rings because of their advantage of having a constant uniform curvature defined by their radius, $q = 1/R$ (**Fig. 2, a**). The BLA pattern is retained in the rings having an even number of atoms, as seen from the electronic density distribution in **Fig. 2 (a)**. Additional care must be taken with carbyne rings since the Jahn–Teller distortion (the counterpart of Peierls instability in non-linear molecules) is different in the $C_{4N}$ and $C_{4N+2}$ families of rings.[52–54] The $C_{4N+2}$ rings undergo a second-order distortion which produces little/no BLA at small $N$. Because of this, the bending stiffness in the two families differs for small rings, but as $N$ increases ($q \to 0$), the difference gradually disappears, and the stiffness of both families approaches the common limit of infinite straight carbyne, which the extrapolation of our results places at $K = 3.56$ **eV·Å**. A similar extrapolation of the ring energies disagrees with an explicit periodic infinite-chain calculation by only about 0.3 meV/atom, showing the robustness of our procedure.

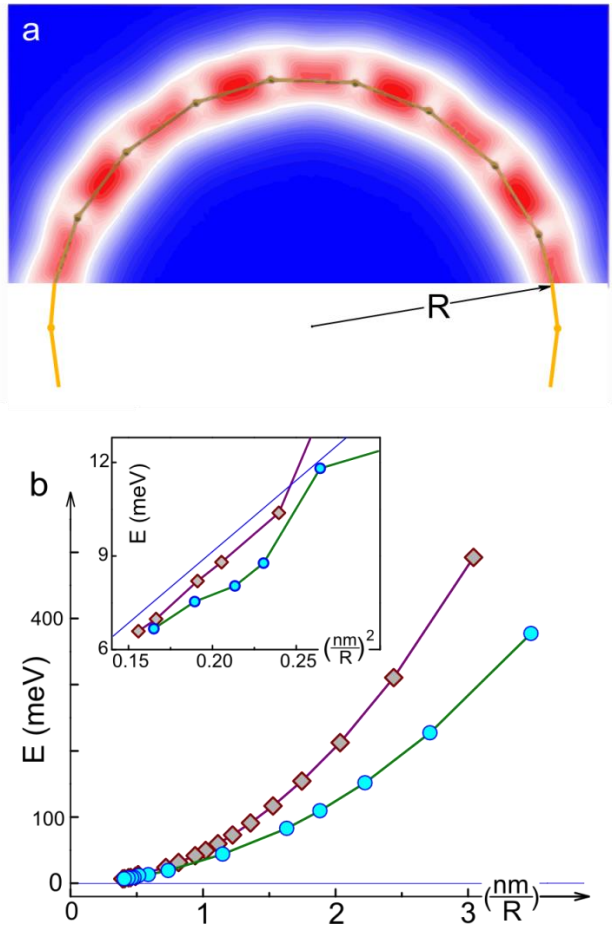

**Figure 2.** Bending stiffness of carbyne. (a) The carbon ring model with its electron density distribution, showing the usual bond alternation pattern. (b) Bending energy per unit cell as a function of ring curvature calculations used to extract the bending modulus value.

**Persistence length.** In order to determine whether this value of $K$ is large or small, we can further characterize the bending stiffness of carbyne using the concept of persistence length familiar from polymer physics. It is defined as $l_p = K/k_B T$, where $k_B$ is the Boltzmann constant and $T$ is the temperature. Using our value of $K$, the persistence length of carbyne at 300 K is roughly **$l_p$ = 14 nm**, or about 110 C atoms. This can be compared to the persistence length of various important polymers, as presented in Table 1. As we see, carbyne is relatively stiff for a polymer and comparable to narrowest graphene nanoribbons, and it is about 13 times as stiff as a dsDNA rod (mind the 2-nm diameter of the latter).

**Table 1.** Persistence length of carbyne as compared to several important polymers

| Polymer | $l_p$ (nm) |
| --- | --- |
| polyacrylonitrile[67] | 0.4–0.6 |
| polyacetylene[68] | 1.3 |
| single-stranded DNA[69,70] | 1–4 |
| polyaniline[71] | 9 |
| double-stranded DNA (dsDNA)[56] | 45–50 |
| graphene nanoribbons[72] | 10–100 |
| **carbyne (present work)** | **14** |

**Carbyne under torsion: chemically modulated spring.** The torsional stiffness of a rod of length $L$ is expressed as

$$H = L \frac{\partial^2 E}{\partial \theta^2}, \qquad (3)$$

where $\theta$ is the torsion angle. Looking at the electron density distribution along the atomic chain (**Fig. 3, a**), one immediately recognizes a problem with this definition. Because of the cylindrical symmetry of the density—and, consequently, all relevant physical observables—it is impossible to define $\theta$, and thus $H$ has to be zero. However, we can solve this problem by attaching functional group 'handles' at the chain ends to break the cylindrical symmetry (**Fig. 3, b** and **c**). Further, depending on the character of the passivation—$sp^3$ or $sp^2$—one should expect different effects on the torsional stiffness.[30]

We considered different functional groups, including methyl ($CH_3$), phenyl ($C_6H_5$), and hydroxyl (OH) as $sp^3$ radicals, and amine ($NH_2$, planar) and methylene ($CH_2$) groups from the $sp^2$ family. With methyl and hydroxyl, we find that the single C–R bond at the end interfaces so smoothly with the single–triple bond system of carbyne as to result in a completely free unhindered rotation. Moreover, phenyl handles also rotated freely, suggesting a single bond between the carbyne and the aromatic ring. Only the =CH$_2$ (and, weakly, NH$_2$) termination shows a detectable torsional stiffness. As the "handles" are twisted out-of-plane, the frontier π orbitals assume a characteristic helical structure with the total twist angle following the dihedral angle between the handle group planes—see **Fig. 3 (b,c)** and **Fig. S2** (while the present paper was under revision, an independent study on helical orbitals in conjugated molecules came out[55]).

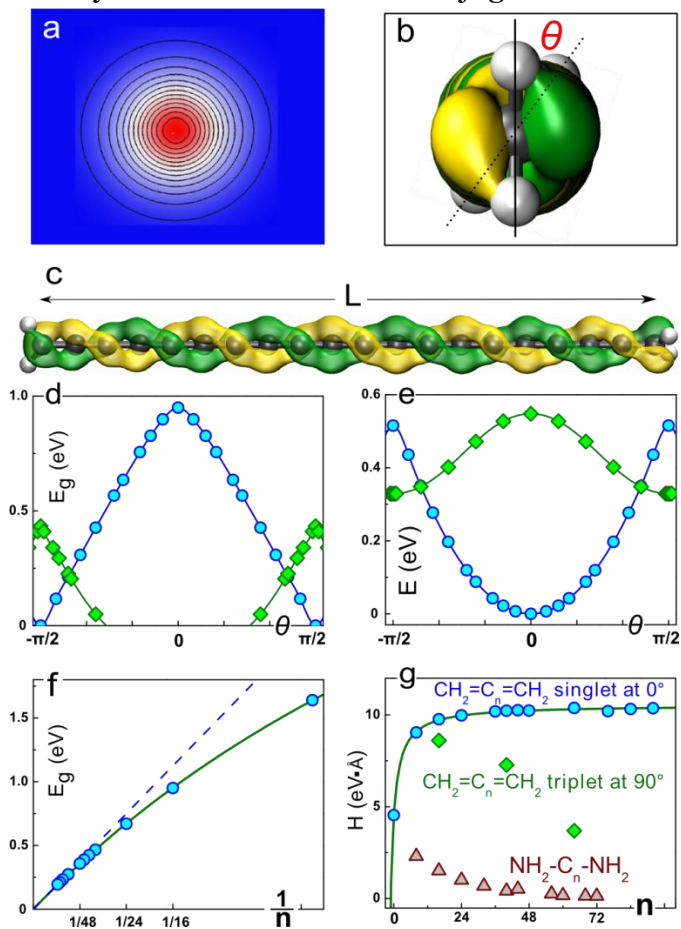

**Figure 3.** End-group-induced torsional stiffness of carbyne. (a) Cross-section of the electronic density of carbyne. (b) Front and (c) side view of a carbyne chain with attached =CH$_2$ handles, showing the helical HOMO isosurface. (d) Band gap and (e) energy as a function of the torsional angle: blue, singlet state; green, triplet. (f) Band gap and (g) torsional stiffness dependence on the chain length. The blue line in (f) is fitted by $E = A/L$ and green line is fitted by $E = A/(L + L_o)$

Before we present the numerical results, an interesting question to consider is how this stiffness could emerge at the electronic-structure level. Consider the behavior of band gap as a function of the torsional angle. As the torsional angle increases, the band gap shrinks and completely closes at 90°, as shown in **Fig. 3 (d)**. From the molecular orbitals perspective, at 0°, the planar geometry can sustain two fully orthogonal *p*-bands formed by $p_y$ and $p_z$ orbitals, respectively,—the HOMO and LUMO of the molecule ($D_{2h}$ symmetry). However, when the handles are at 90° ($D_{2d}$), the two orbitals become completely equivalent by symmetry and, hence, degenerate. As a result, the energy of the system is increased—**Fig. 3 (e)**. We can estimate this energy change as $2 \times [E_g/2] = E_g$, the band gap of the planar molecule, and the resulting potential energy (PE) curve can be approximated by its first Fourier term as $E(\theta) = \frac{1}{2}(1 - \cos 2\theta)E_g$. Then, according to Eq. (3), the torsional stiffness is $H(E_g, L) = 2E_g \times L$.

Next we examine the behavior of $E_g$ as a function of $L$. Treating carbyne as a 1D potential well of width $L \equiv Na$, the HOMO and LUMO energies are $E_N \propto k_N^2 \propto N^2/L^2$ and $E_{N+1} \propto k_{N+1}^2 \propto (N+1)^2/L^2$, and therefore, $E_g \propto (1+2N)/L^2 \propto 1/L$ for long chains, $N \gg 1$. The inverse dependence of the band gap on the length, $E_g = C/L$, leads to $H = 2C = const$, i.e., the torsional stiffness of carbyne has a finite constant value.

Both predictions—$E_g \propto 1/L$ and a constant torsional stiffness $H$—are confirmed by our DFT calculations, as shown in **Fig. 3 (f)** and **(g)**, respectively. Although this stiffness is not, strictly speaking, inherent to carbyne in the sense that it relies on the endgroups to break the rotational symmetry, from the mechanical perspective the carbon chain between the handles behaves just like a classical torsion rod. With the =CH$_2$ termination, the maximum stiffness value we observe is $H \approx$ **10.3 eV·Å/rad²**. This figure is about half the torsional stiffness of dsDNA,[56] which is truly remarkable given carbyne's one-atom thickness.

Two more interesting aspects of carbyne's behavior as a torsional spring deserve to be highlighted. First, upon passing 90° twist, the energy evolution is reversed, and the differential stiffness becomes formally negative. This behavior is quite intriguing, since for a normal beam, the energy would increase monotonously during twisting, whereas for this 1-D carbon chain, the torsional energy exhibits a periodic behavior. Second, as is typical for such situations, the frontier orbital degeneracy at 90° can be lifted *via* spin unpairing. Indeed, our calculations show that at near-90° angles, the spin-triplet state is energetically more favorable[30] and, in fact, becomes the local energy minimum (**Fig. 3 (e)**). In this state carbyne essentially becomes a magnetic semiconductor, hinting at applications in spintronics. A separate torsional stiffness can be defined for this minimum, $H_T$ = 8.6 and 3.7 eV·Å/rad² for $N$ = 18 and 66 carbyne chains, respectively.

**Equivalent continuum-mechanics model.** To gain further insight and appreciation of the mechanical properties of carbyne, we need to establish the link between the computed atomistic properties and the concepts of continuum elasticity such as the Young's modulus $Y$, the bending stiffness $K$, the shear modulus $G$. The fully assembled system of equations looks, where all the atomistic quantities (Eqs. 1–3) are assembled on the left and their counterparts from continuum mechanics, on the right.

$$C = \frac{1}{L^2}\frac{\partial^2 E}{\partial \varepsilon^2} = \frac{\pi r^2 Y}{L}, \tag{4}$$

$$K = \frac{1}{L}\frac{\partial^2 E}{\partial (1/R)^2} = \frac{1}{4}\pi r^4 Y, \tag{5}$$

$$H = L\frac{\partial^2 E}{\partial \theta^2} = 2G\pi r^4. \tag{6}$$

From Eqs. (4) and (5), we can estimate the effective mechanical radius and the Young's modulus of carbyne: **$r$ = 0.386 Å** and **$Y$ = 32.71 TPa**. Knowing the radius, we are now able to compute the shear modulus from (6): **$G$ = 11.8 TPa**, and thus the effective Poisson's ratio $n = Y/(2G) - 1$, **$n$ = 0.386**. Comparing with the hardest natural materials diamond and graphene ($Y_{diamond}$ = 1.22 TPa, $Y_{graphene}$ = 1 TPa, $G_{diamond}$ = 0.5 TPa), we see that carbyne, outperforms them by a factor of about 30 (owing in part to its extremely small mechanical radius). For convenience, the elastic parameters of carbyne are summarized in Table 2.

**Table 2.** Summary of mechanical characteristics of carbyne

| Property | Value |
| --- | --- |
| thickness 2r | 0.772 Å |
| Young's modulus Y | 32.71 TPa |
| shear modulus G | 11.8 TPa |
| Poisson's ratio n | 0.386 |
| persistence length at 300 K | 14 nm |

**Carbyne as a conducting cable.** To investigate the characteristics of carbyne as a conductor, we computed the distribution of electrical current density using the standard real-space nonequilibrium Green's function approach. It gives an explicit form of current density at a given point in space:[57]

$$J(\mathbf{r}) = \frac{1}{2}\sum_{i,j}\int dE \cdot$$
$$\cdot G_{ij}^{<}(E) \lim_{r' \to r}(\nabla' - \nabla)\,\phi_i(\mathbf{r}')\phi_j^*(\mathbf{r}), \tag{7}$$

where $\phi_i$ are the atomic basis functions (2s and 2p), $G^{<}(E)$ is the matrix correlation function related to the electron density by integration over the energy, and the gradient part corresponds to the probability flux. In addition to integrating the transmission coefficients over the energy range between the two Fermi levels, the total current can also be retrieved by integrating the current density over an arbitrary cross-section of the carbon chain.

**Fig. 4 (a)** shows the calculated current density distribution of a carbon chain. The distribution is axially-symmetric with negligible magnitudes on the axis and a bulge at about $r = 1$ Å. This reflects the fact that the $\sigma$ bonds are localized between atoms and form the skeleton of carbon structures, while the delocalized $\pi$ bonds normal to the axis support the electronic conductance. By the Biot–Savart–Laplace law, the induced magnetic field can also be determined. As shown in **Fig. 4 (b)**, it exhibits a similar 'tubular' distribution pattern.

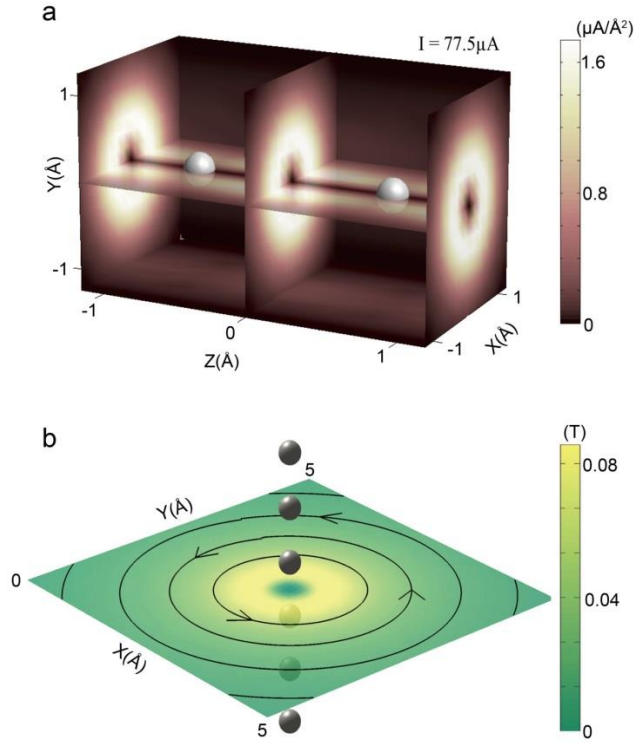

**Figure 4.** Carbyne as a conducting cable. (a) Current density distributions in the axial plane of carbyne and in cross-section. (b) The magnetic field induced by the current density of (a). The directions of magnetic vectors correspond to the current flowing from the bottom up.

Similarly to the previous section, it is instructive to estimate what would be the equivalent of carbyne in classical conductors. We may define the effective radius $r_{eff}$ of the carbyne as the radius of a conducting cylinder that generates an identical inductance to that of carbyne at the same electrical current. The magnetic flux of the classical conductor per unit length is evaluated by

$$\Phi = \frac{\mu_0}{2\pi}\left(\frac{1}{2} + \ln \frac{R}{r_0}\right)I, \tag{8}$$

where $R$ is the length of the flux surface, $r_0$ is the cylinder radius, and $I$ is the current. With the magnetic field obtained above, the magnetic flux of the carbyne carrying a current $I$ can be computed if $R$ is given. We observe a linear dependence between $\Phi$ and $I$ of carbyne for various values of $R = 10–100$ Å. Although the slope varies as $R$ changes, each of them leads to the same $r_0$ by Eq. (8). Accordingly, the effective radius of carbyne in terms of inductance is found to be $\boldsymbol{r_{eff} = 1.196}$ Å.

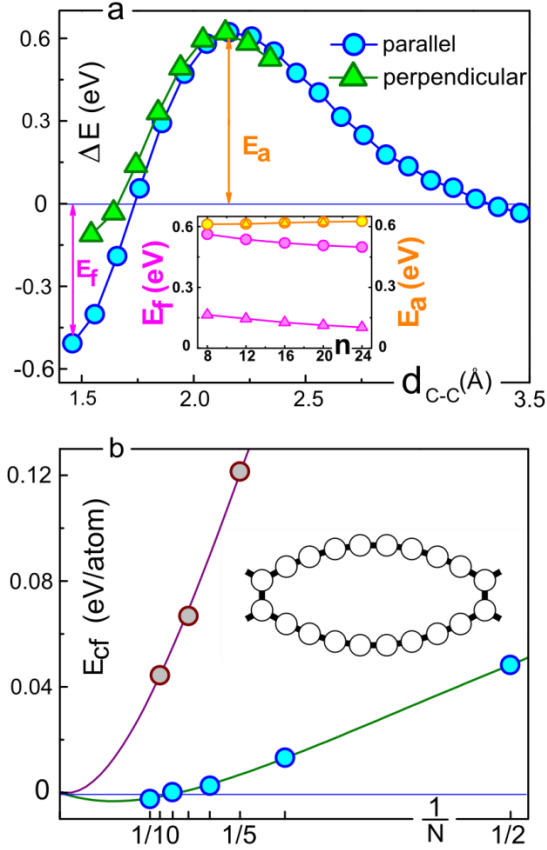

**Figure 5.** Cross-linking of carbyne chains. (a) Potential energy scans showing the barrier of ~0.6 eV for finite-length chain crosslinking. (b) Energy per atom in periodically cross-linked systems with $N$ atoms between cross-links, showing a maximum energetically favorable cross-link density of 1 link per 9–11 atoms, and an equilibrium spacing of 16 atoms. $E = 0$ corresponds to the reference energy level of infinitely isolated carbyne chains.

**Chemical stability of carbyne.** An important concern for any nanoscale material is its stability with respect to dimerization.[58] For carbyne, formation of cross-links between chains is a mechanism of degradation known from experiments,[11] and hence must be studied. **Fig. 5 (a)** shows a relaxed potential energy surface scan using the distance between C atoms in the middle of two carbyne chains as the constraint onto which the reaction coordinate is projected. To avoid introducing bias due to the cumulene/polyyne transformation during the cross-link formation, we used two chains with $2N$ and $2N + 1$ atoms to simulate the reaction, so that both before and after the reaction there is exactly one chain with an (unfavorable) odd number of atoms in the system. We estimated an activation energy barrier of about $E_a$ = **0.6 eV** that is independent of the length of reacting chains (from 8 to 24 atoms) as well as of the dihedral angle formed by the chains. (The overall exothermicity of the reaction showed a weak decrease with the chain length; *i.e.*, shorter chains have a stronger tendency to link.) This barrier suggests the viability of carbyne in condensed phase at room temperature on the order of days.[58]

What will happen if two parallel long carbyne chains are brought into contact? Every cross-link formed will create local curvature, raising the energy of the system. We can estimate the maximum frequency of such cross-links using a simple geometrical model. Assuming all bond angles about a cross-link being 120° and the carbyne curvature between the links remains constant, the length of the arc between two cross-links is $L = (\pi/3)R$. The corresponding elastic energy is $E_{bend} = KL/2R^2 = K\pi^2/18L = 1.95/L$ eV. This energy cost is offset by the gain from forming the extra bonds between the chains, which is about $E_f$ = 0.25 eV/atom from **Fig. 5 (a)**. Using this value, we arrive at an estimate for $L \approx$ 7.8 Å, or about $N$ = 5 atoms between the cross-links. Compared to the results of DFT calculations presented in **Fig. 5 (b)**, which predict an 8-atom threshold spacing, this is an underestimate that can be straightforwardly attributed to our model's ignoring the π-electron conjugation which is perturbed by the cross-links. Since the energy of conjugation interruption should also scale approximately as $1/L$, the general form of $E_{bend+\pi} = K^*/L$ should still hold. In this case, it is easy to show that the *equilibrium* separation between kinks in an infinite system (one that minimizes the total energy per atom) is exactly twice the *threshold* separation (at which $E_{bend+\pi}$ = $E_f$), and can be estimated from our DFT data as one cross-link per 17 atoms, or 21.8 Å.

Note the oscillating behavior of the DFT data points as the number of atoms between sequential cross-links switches between odd and even. It indicates that the atoms that make the cross-link form a double bond and thus are connected to the chain segments by single bonds, which is only possible to seamlessly integrate into the single–triple alternation pattern of polyyne when the number of atoms in between is even. In fact, odd-number systems are instable in the sense that if the unit cell is doubled, an $N$ = $2M$+1 system tends to reconstruct into a {$2P$, $2(M - P)$} superstructure. An alternative that is favorable for shorter-period systems is to form a pair of adjacent cross-links, resulting in a square $C_4$ unit (see Supplementary Information for details).

Our finding of a large (~2 nm) equilibrium cross-link spacing explains the experimental observations where single-zigzag–wide graphene nanoribbons show instability and split into pairs of atomic chains.[14] In summary, carbyne chains can be robust against cross-linking at not-too-high temperatures, and mechanical constraints can further prevent them from forming multiple bonds with each other.

## Summary and conclusions

The comprehensive 'portrait' of carbyne that we have drawn can be formulated like this. It has an extreme tensile stiffness—stiffer by a factor of two than graphene and carbon nanotubes—and a specific strength surpassing that of any other known material. Its flexibility is between those of typical polymers and double-stranded DNA, with a persistence length of ~14 nm. It is equivalent to a continuum-mechanics rod of an extremely small diameter, 0.772 Å, and an enormous nominal Young's modulus of 32.7 TPa. Carbyne can be toggled from the free-rotating to torsionally stiff state by appropriate chemical functionalization (=CH$_2$), in which case its effective shear modulus is about 11.8 TPa, and its formal Poisson's ratio is 0.386. With this or similar functionalization, a magnetic-semiconductor behavior can be switched on by a ~90° twist. Its band gap increases under tension from 2.6 up to 4.7 eV at a 10% strain, and it is reasonably stable chemically, with an activation barrier of 0.6 eV for cross-linking and an equilibrium link density of one per 2.2 nm (17 atoms) due to $\pi$-electron confinement. This combination of unusual mechanical and electronic properties is of great interest for applications in nanomechanical systems, opto-/electromechanical devices, strong and light materials for mechanical applications, or as high–specific-area energy storage matrices.

## Methods

The density functional theory and GW calculations were performed with VASP[59,60]. Mechanical properties were computed using the PBE[61,62] generalized-gradient exchange–correlation functional and a 400 eV plane wave cutoff. 12-Å cell spacing was used in the aperiodic directions to eliminate the interaction of periodic images of the system. Structural relaxation was performed until all forces were less than 0.01 eV/Å. Chemical reaction barrier was calculated using the CEP-121G* basis set[63] in Gaussian 09[64] (spin-unrestricted). The Hamiltonian and overlap matrix elements for quantum transport calculations were computed using SIESTA.[65,66] More detail is available in SI.

**Supporting Information**. Details of tensile strength, torsional stiffness, and chemical stability calculations. This material is available free of charge *via* the Internet at http://pubs.acs.org.

AUTHOR INFORMATION
**Corresponding Author**
* e-mail: biy@rice.edu

**Present Addresses**
† Department of Physics, Konkuk University, Seoul, 143-701, Korea

ACKNOWLEDGMENT

We thank A. Kutana and X. Zou for assistance with GW calculations. The research was supported by the Robert Welch Foundation (C-1590) and the Department of Energy. The computational resources were provided under NSF support.

REFERENCES

1. Kroto, H. W.; Heath, J. R.; O'Brien, S. C.; Curl, R. F.; Smalley, R. E. $C_{60}$: Buckminsterfullerene. *Nature* **1985**, *318*, 162–163.

2. Boehm, H. P.; Clauss, A.; Fischer, G.; Hofmann, U. Thin Carbon Leaves. *Z. Für Naturforschung* **1962**, *17b*, 150–153.

3. Novoselov, K. S.; Geim, A. K.; Morozov, S. V.; Jiang, D.; Zhang, Y.; Dubonos, S. V.; Grigorieva, I. V.; Firsov, A. A. Electric Field Effect in Atomically Thin Carbon Films. *Science* **2004**, *306*, 666–669.

4. Bolotin, K.; Sikes, K.; Jiang, Z.; Klima, M.; Fudenberg, G.; Hone, J.; Kim, P.; Stormer, H. Ultrahigh Electron Mobility in Suspended Graphene. *Solid State Commun.* **2008**, *146*, 351–355.

5. Whittaker, A. G. Carbon: A New View of Its High-Temperature Behavior. *Science* **1978**, *200*, 763–764.

6. Goresy, A. E.; Donnay, G. A New Allotropic Form of Carbon from the Ries Crater. *Science* **1968**, *161*, 363–364.

7. Webster, A. Carbyne as a Possible Constituent of the Interstellar Dust. *Mon. Not. R. Astron. Soc.* **1980**, *192*, 7–9.

8. Hayatsu, R.; Scott, R. G.; Studier, M. H.; Lewis, R. S.; Anders, E. Carbynes in Meteorites: Detection, Low-Temperature Origin, and Implications for Interstellar Molecules. *Science* **1980**, *209*, 1515–1518.

9. Jones, R. O.; Seifert, G. Structure and Bonding in Carbon Clusters $C_{14}$ to $C_{24}$: Chains, Rings, Bowls, Plates, and Cages. *Phys. Rev. Lett.* **1997**, *79*, 443–446.

10. Lagow, R. J.; Kampa, J. J.; Wei, H.-C.; Battle, S. L.; Genge, J. W.; Laude, D. A.; Harper, C. J.; Bau, R.; Stevens, R. C.; Haw, J. F.; *et al.* Synthesis of Linear Acetylenic Carbon: The "*sp*" Carbon Allotrope. *Science* **1995**, *267*, 362–367.

11. Kavan, L.; Hlavatý, J.; Kastner, J.; Kuzmany, H. Electrochemical Carbyne from Perfluorinated Hydrocarbons: Synthesis and Stability Studied by Raman Scattering. *Carbon* **1995**, *33*, 1321–1329.

12. Casari, C. S.; Bassi, A. Li; Ravagnan, L.; Siviero, F.; Lenardi, C.; Piseri, P.; Bongiorno, G.; Bottani, C. E.; Milani, P. Chemical and Thermal Stability of Carbyne-Like Structures in Cluster-Assembled Carbon Films. *Phys. Rev. B* **2004**, *69*, 075422.

13. *Polyynes : Synthesis, Properties, and Applications*; Cataldo, F., Ed.; CRC Press: Boca Raton, 2005.

14. Jin, C.; Lan, H.; Peng, L.; Suenaga, K.; Iijima, S. Deriving Carbon Atomic Chains from Graphene. *Phys. Rev. Lett.* **2009**, *102*, 205501.

15. Hobi, E.; Pontes, R. B.; Fazzio, A.; da Silva, A. J. R. Formation of Atomic Carbon Chains from Graphene Nanoribbons. *Phys. Rev. B* **2010**, *81*, 201406.

16. Erdogan, E.; Popov, I.; Rocha, C. G.; Cuniberti, G.; Roche, S.; Seifert, G. Engineering Carbon Chains from Mechanically Stretched Graphene-Based Materials. *Phys. Rev. B* **2011**, *83*, 041401.

17. Yakobson, B. I.; Campbell, M. P.; Brabec, C. J.; Bernholc, J. High Strain Rate Fracture and C-Chain Unraveling in Carbon Nanotubes. *Comput. Mater. Sci.* **1997**, *8*, 341–348.

18. Chalifoux, W.; Tykwinski, R. Synthesis of Polyynes to Model the *sp*-Carbon Allotrope Carbyne. *Nat. Chem.* **2010**, *2*, 967–971.

19. Khoo, K. H.; Neaton, J. B.; Son, Y. W.; Cohen, M. L.; Louie, S. G. Negative Differential Resistance in Carbon Atomic Wire-Carbon Nanotube Junctions. *Nano Lett.* **2008**, *8*, 2900–2905.

20. Zanolli, Z.; Onida, G.; Charlier, J.-C. Quantum Spin Transport in Carbon Chains. *ACS Nano* **2010**, *4*, 5174–5180.

21. Zeng, M. G.; Shen, L.; Cai, Y. Q.; Sha, Z. D.; Feng, Y. P. Perfect Spin-Filter and Spin-Valve in Carbon Atomic Chains. *Appl. Phys. Lett.* **2010**, *96*, 042104–042104–3.

22. Akdim, B.; Pachter, R. Switching Behavior of Carbon Chains Bridging Graphene Nanoribbons: Effects of Uniaxial Strain. *ACS Nano* **2011**, *5*, 1769–1774.

23. Artyukhov, V. I.; Liu, M.; Yakobson, B. I. Mechanically Induced Metal-Insulator Transition in Carbyne. *arXiv:1302.7250* **2013**.

24. Sorokin, P. B.; Lee, H.; Antipina, L. Y.; Singh, A. K.; Yakobson, B. I. Calcium-Decorated Carbyne Networks as Hydrogen Storage Media. *Nano Lett.* **2011**, *11*, 2660–2665.

25. Movsisyan, L. D.; Kondratuk, D. V.; Franz, M.; Thompson, A. L.; Tykwinski, R. R.; Anderson, H. L. Synthesis of Polyyne Rotaxanes. *Organics Lett.* **2012**, *14*, 3424–3426.

26. Weisbach, N.; Baranová, Z.; Gauthier, S.; Reibenspies, J. H.; Gladysz, J. A. A New Type of Insulated Molecular Wire: a Rotaxane Derived from a Metal-Capped Conjugated Tetrayne. *Chem. Commun.* **2012**, *48*, 7562–7564.

27. Liu, M.; Lee, H.; Yakobson, B. I. Carbynes by the First Principles Computations: an Ultimate 1D-Material. In *ASME 2012 International Mechanical Engineering Congress & Exposition*; Houston, TX, USA, 2012; session 10-6-1; abs. IMECE2012–89610.

28. Abdurahman, A.; Shukla, A.; Dolg, M. Ab Initio Many-Body Calculations on Infinite Carbon and Boron-Nitrogen Chains. *Phys. Rev. B* **2002**, *65*, 115106.

29. Tongay, S.; Durgun, E.; Ciraci, S. Atomic Strings of Group IV, III–V, and II–VI Elements. *Appl. Phys. Lett.* **2004**, *85*, 6179–6181.

30. Ravagnan, L.; Manini, N.; Cinquanta, E.; Onida, G.; Sangalli, D.; Motta, C.; Devetta, M.; Bordoni, A.; Piseri, P.; Milani, P. Effect of Axial Torsion on *sp* Carbon Atomic Wires. *Phys. Rev. Lett.* **2009**, *102*, 245502.

31. Hu, Y. H. Bending Effect of *sp*-Hybridized Carbon (Carbyne) Chains on Their Structures and Properties. *J. Phys. Chem. C* **2011**, *115*, 1843–1850.

32. Nair, A. K.; Cranford, S. W.; Buehler, M. J. The Minimal Nanowire: Mechanical Properties of Carbyne. *EPL Eur. Lett.* **2011**, *95*, 16002.

33. Castelli, I. E.; Salvestrini, P.; Manini, N. Mechanical Properties of Carbynes Investigated by Ab Initio Total-Energy Calculations. *Phys. Rev. B* **2012**, *85*, 214110.

34. Kertesz, M.; Koller, J.; Azman, A. Ab Initio Hartree–Fock Crystal Orbital Studies. II. Energy Bands of an Infinite Carbon Chain. *J. Chem. Phys.* **1978**, *68*, 2779–2782.

35. Pomerantz, A. E.; Han, J. H.; Musgrave, C. B. Calculating Cumulene/Poly-yne Isomerization Energies. *J. Phys. Chem. A* **2004**, *108*, 4030–4035.

36. Tongay, S.; Senger, R. T.; Dag, S.; Ciraci, S. Ab-Initio Electron Transport Calculations of Carbon Based String Structures. *Phys. Rev. Lett.* **2004**, *93*, 136404.

37. Zhao, X.; Ando, Y.; Liu, Y.; Jinno, M.; Suzuki, T. Carbon Nanowire Made of a Long Linear Carbon Chain Inserted Inside a Multiwalled Carbon Nanotube. *Phys. Rev. Lett.* **2003**, *90*, 187401.

38. Lee, C.; Wei, X.; Kysar, J. W.; Hone, J. Measurement of the Elastic Properties and Intrinsic Strength of Monolayer Graphene. *Science* **2008**, *321*, 385–388.

39. Liu, F.; Ming, P.; Li, J. Ab Initio Calculation of Ideal Strength and Phonon Instability of Graphene under Tension. *Phys. Rev. B* **2007**, *76*, 064120.

40. Sánchez-Portal, D.; Artacho, E.; Soler, J. M.; Rubio, A.; Ordejón, P. Ab Initio Structural, Elastic, and Vibrational Properties of Carbon Nanotubes. *Phys. Rev. B* **1999**, *59*, 12678–12688.

41. Spear, K. E.; Dismukes, J. P. *Synthetic Diamond: Emerging CVD Science and Technology*; Electrochemical Society; Wiley: New York, 1994.

42. Cahangirov, S.; Topsakal, M.; Ciraci, S. Long-Range Interactions in Carbon Atomic Chains. *Phys. Rev. B* **2010**, *82*, 195444.

43. Dumitrica, T.; Hua, M.; Yakobson, B. I. Symmetry-, Time-, and Temperature-Dependent Strength of Carbon Nanotubes. *Proc. Natl. Acad. Sci.* **2006**, *103*, 6105–6109.

44. Ogata, S.; Shibutani, Y. Ideal Tensile Strength and Band Gap of Single-Walled Carbon Nanotubes. *Phys. Rev. B* **2003**, *68*, 165409.

45. Chang, C.-C.; Hsu, I.-K.; Aykol, M.; Hung, W.-H.; Chen, C.-C.; Cronin, S. B. A New Lower Limit for the Ultimate Breaking Strain of Carbon Nanotubes. *ACS Nano* **2010**, *4*, 5095–5100.

46. Telling, R. H.; Pickard, C. J.; Payne, M. C.; Field, J. E. Theoretical Strength and Cleavage of Diamond. *Phys. Rev. Lett.* **2000**, *84*, 5160–5163.

47. Jacquemin, D.; Femenias, A.; Chermette, H.; Ciofini, I.; Adamo, C.; André, J.-M.; Perpète, E. A. Assessment of Several Hybrid DFT Functionals for the Evaluation of Bond Length Alternation of Increasingly Long Oligomers. *J. Phys. Chem. A* **2006**, *110*, 5952–5959.

48. Heyd, J.; Scuseria, G. E.; Ernzerhof, M. Hybrid Functionals Based on a Screened Coulomb Potential. *J. Chem. Phys.* **2003**, *118*, 8207–8215.

49. Heyd, J.; Scuseria, G. E.; Ernzerhof, M. Erratum: "Hybrid Functionals Based on a Screened Coulomb Potential" [ J. Chem. Phys. 118, 8207 (2003) ]. *J. Chem. Phys.* **2006**, *124*, 219906–219906–1.

50. Hybertsen, M. S.; Louie, S. G. Electron Correlation in Semiconductors and Insulators: Band Gaps and Quasiparticle Energies. *Phys. Rev. B* **1986**, *34*, 5390–5413.

51. Cretu, O.; Botello-Mendez, A. R.; Janowska, I.; Pham-Huu, C.; Charlier, J.-C.; Banhart, F. Electrical Transport Measured in Atomic Carbon Chains. *Nano Lett.* **2013**, *13*, 3487–3493.

52. Bylaska, E. J.; Weare, J. H.; Kawai, R. Development of Bond-Length Alternation in Very Large Carbon Rings: LDA Pseudopotential Results. *Phys. Rev. B* **1998**, *58*, R7488–R7491.

53. Torelli, T.; Mitas, L. Electron Correlation in $C_{4N+2}$ Carbon Rings: Aromatic *Versus* Dimerized Structures. *Phys. Rev. Lett.* **2000**, *85*, 1702–1705.

54. Saito, M.; Okamoto, Y. Second-Order Jahn–Teller Effect on Carbon 4*N*+2 Member Ring Clusters. *Phys. Rev. B* **1999**, *60*, 8939–8942.

55. Hendon, C. H.; Tiana, D.; Murray, A. T.; Carbery, D. R.; Walsh, A. Helical Frontier Orbitals of Conjugated Linear Molecules. *Chem. Sci.* **2013**, *4*, 4278–4284.

56. Hagerman, P. J. Flexibility of DNA. *Annu. Rev. Biophys. Biophys. Chem.* **1988**, *17*, 265–286.

57. Xue, Y.; Datta, S.; Ratner, M. A. First-Principles Based Matrix Green's Function Approach to Molecular Electronic Devices: General Formalism. *Chem. Phys.* **2002**, *281*, 151–170.

58. Hoffmann, R. Small but Strong Lessons from Chemistry for Nanoscience. *Angew. Chem. Int. Ed.* **2013**, *52*, 93–103.

59. Kresse, G.; Hafner, J. Ab Initio Molecular Dynamics for Liquid Metals. *Phys. Rev. B* **1993**, *47*, 558–561.

60. Kresse, G.; Furthmüller, J. Efficient Iterative Schemes for Ab Initio Total-Energy Calculations Using a Plane-Wave Basis Set. *Phys. Rev. B* **1996**, *54*, 11169–11186.

61. Perdew, J. P.; Burke, K.; Ernzerhof, M. Generalized Gradient Approximation Made Simple. *Phys. Rev. Lett.* **1996**, *77*, 3865–3868.

62. Perdew, J. P.; Burke, K.; Ernzerhof, M. Generalized Gradient Approximation Made Simple [Phys. Rev. Lett. 77, 3865 (1996)]. *Phys. Rev. Lett.* **1997**, *78*, 1396.

63. Stevens, W. J.; Basch, H.; Krauss, M. Compact Effective Potentials and Efficient Shared-exponent Basis Sets for the First- and Second-Row Atoms. *J. Chem. Phys.* **1984**, *81*, 6026–6033.

64. Frisch, M. J.; Trucks, G. W.; Schlegel, H. B.; Scuseria, G. E.; Robb, M. A.; Cheeseman, J. R.; Scalmani, G.; Barone, V.; Mennucci, B.; Petersson, G. A.; *et al. Gaussian 09 Revision A.1*.

65. Soler, J. M.; Artacho, E.; Gale, J. D.; García, A.; Junquera, J.; Ordejón, P.; Sánchez-Portal, D. The SIESTA Method for Ab Initio Order-N Materials Simulation. *J. Phys. Condens. Matter* **2002**, *14*, 2745–2779.

66. Artacho, E.; Anglada, E.; Diéguez, O.; Gale, J. D.; García, A.; Junquera, J.; Martin, R. M.; Ordejón, P.; Pruneda, J. M.; Sánchez-Portal, D.; *et al.* The SIESTA Method; Developments and Applicability. *J. Phys. Condens. Matter* **2008**, *20*, 064208.

67. Ganster, J.; Fink, H.-P.; Zenke, I. Chain Conformation of Polyacrylonitrile: a Comparison of Model Scattering and Radial Distribution Functions with Experimental Wide-Angle X-Ray Scattering Results. *Polymer* **1991**, *32*, 1566–1573.

68. Severin, M.; Inganäs, O. Global Conformation and Order Parameters of Doped Polyacetylene at Finite Temperatures. *Eur. Lett. EPL* **1994**, *25*, 347–352.

69. Smith, S. B.; Cui, Y.; Bustamante, C. Overstretching B-DNA: The Elastic Response of Individual Double-Stranded and Single-Stranded DNA Molecules. *Science* **1996**, *271*, 795–799.

70. Tinland, B.; Pluen, A.; Sturm, J.; Weill, G. Persistence Length of Single-Stranded DNA. *Macromolecules* **1997**, *30*, 5763–5765.

71. Gettinger, C. L.; Heeger, A. J.; Pine, D. J.; Cao, Y. Solution Characterization of Surfactant Solubilized Polyaniline. *Synth. Met.* **1995**, *74*, 81–88.

72. Bets, K. V.; Yakobson, B. I. Spontaneous Twist and Intrinsic Instabilities of Pristine Graphene Nanoribbons. *Nano Res* **2009**, *2*, 161–166.